\documentclass[article]{jss}

\author{Yaohui Zeng\\University of Iowa \And 
        Patrick Breheny\\University of Iowa}
\title{The \pkg{biglasso} Package: A Memory- and Computation-Efficient Solver for Lasso Model Fitting with Big Data in \proglang{R}}

\Plainauthor{Yaohui Zeng, Patrick Breheny} %% comma-separated
\Plaintitle{biglasso Package} %% without formatting
\Shorttitle{\pkg{biglasso}: A Memory- and Computation-Efficient \proglang{R} package for Lasso} %% a short title (if necessary)

\Abstract{
  Penalized regression models such as the lasso have been extensively applied to analyzing high-dimensional data sets. However, due to memory limitations, existing \proglang{R} packages like \pkg{glmnet} and \pkg{ncvreg} are not capable of fitting lasso-type models for ultrahigh-dimensional, multi-gigabyte data sets that are increasingly seen in many areas such as genetics, genomics, biomedical imaging, and high-frequency finance. In this research, we implement an \proglang{R} package called \pkg{biglasso} that tackles this challenge. \pkg{biglasso} utilizes memory-mapped files to store the massive data on the disk, only reading data into memory when necessary during model fitting, and is thus able to handle out-of-core computation seamlessly. Moreover, it's equipped with newly proposed, more efficient feature screening rules, which substantially accelerate the computation. Benchmarking experiments show that our \pkg{biglasso} package, as compared to existing popular ones like \pkg{glmnet}, is much more memory- and computation-efficient. We further analyze a 31 GB real data set on a laptop with only 16 GB RAM to demonstrate the out-of-core computation capability of \pkg{biglasso} in analyzing massive data sets that cannot be accommodated by existing \proglang{R} packages.
}
\Plainkeywords{lasso, big data, memory-mapping, out-of-core, hybrid safe-strong rules, parallel computing, C++} %% without formatting
\Address{
  Yaohui Zeng\\
  Department of Biostatistics\\
  University of Iowa\\
  N301 CPHB 145 North Riverside Drive\\
  Iowa City, IA 52242, United States of America\\
  Email: \email{yaohui.zeng@gmail.com}\\

  Patrick Breheny\\
  Department of Biostatistics\\
  University of Iowa\\
  N336 CPHB 145 North Riverside Drive\\
  Iowa City, IA 52242, United States of America\\
  E-mail: \email{patrick-breheny@uiowa.edu}\\
  URL: \url{http://myweb.uiowa.edu/pbreheny/index.html}
}
\usepackage{amsmath, mathrsfs, amsfonts, amssymb, environ}
\usepackage{enumerate}
\usepackage{float}
\usepackage{hyperref}
\usepackage{url}	  % webpage
\usepackage{caption}
\usepackage{multirow}
\usepackage{booktabs} % table ruler
\usepackage{multirow}
\usepackage{bm}
\usepackage{graphicx}
\usepackage{xcolor}
\usepackage{array}
\usepackage{stfloats} % place full-width table at bottom
\usepackage{threeparttable}
\usepackage{adjustbox} % adjust table size
\usepackage{subcaption} % subfigures

\begin{document}
%% Roman letters
\renewcommand{\a}{\mathbf{a}}
\providecommand{\A}{\mathbf{A}}
\renewcommand{\b}{\mathbf{b}}
\providecommand{\B}{\mathbf{B}}
\renewcommand{\c}{\mathbf{c}}
\providecommand{\C}{\mathbf{C}}
\renewcommand{\d}{\mathbf{d}}
\providecommand{\D}{\mathbf{D}}
\providecommand{\e}{\mathbf{e}}
\providecommand{\f}{\mathbf{f}}
\providecommand{\F}{\mathbf{F}}
\renewcommand{\H}{\mathbf{H}}
\providecommand{\I}{\mathbf{I}}
\renewcommand{\j}{\mathbf{j}}
\providecommand{\J}{\mathbf{J}}
\renewcommand{\l}{\mathbf{l}}
\renewcommand{\L}{\mathbf{L}}
\providecommand{\m}{\mathbf{m}}
\providecommand{\M}{\mathbf{M}}
\providecommand{\n}{\mathbf{n}}
\providecommand{\N}{\mathbf{N}}
\providecommand{\q}{\mathbf{q}}
\renewcommand{\P}{\mathbf{P}}
\providecommand{\Q}{\mathbf{Q}}
\renewcommand{\r}{\mathbf{r}}
\providecommand{\bR}{\mathbf{R}}
\providecommand{\s}{\mathbf{s}}
\renewcommand{\S}{\mathbf{S}}
\renewcommand{\u}{\mathbf{u}}
\providecommand{\U}{\mathbf{U}}
\let\vaccent=\v
\renewcommand{\v}{\mathbf{v}}
\providecommand{\V}{\mathbf{V}}
\providecommand{\w}{\mathbf{w}}
\providecommand{\W}{\mathbf{W}}
\providecommand{\x}{\mathbf{x}}
\providecommand{\X}{\mathbf{X}}
\providecommand{\y}{\mathbf{y}}
\providecommand{\z}{\mathbf{z}}
\providecommand{\Z}{\mathbf{Z}}
\providecommand{\one}{\mathbf{1}}
\providecommand{\zero}{\mathbf{0}}

%% Greek letters
\providecommand{\lam}{\lambda}
\providecommand{\Lam}{\Lambda}
\providecommand{\eps}{\epsilon}
\providecommand{\ah}{\hat{\alpha}}
\providecommand{\ba}{\boldsymbol{\alpha}}
\providecommand{\bah}{\hat{\boldsymbol{\alpha}}}
\providecommand{\bb}{\boldsymbol{\beta}}
\providecommand{\bh}{\widehat{\beta}}
\providecommand{\bbh}{\widehat{\boldsymbol{\beta}}}
\providecommand{\bd}{\boldsymbol{\delta}}
\providecommand{\be}{\boldsymbol{\eta}}
\providecommand{\eh}{\hat{\eta}}
\providecommand{\beh}{\hat{\boldsymbol{\eta}}}
\providecommand{\bep}{\boldsymbol{\epsilon}}
\providecommand{\eph}{\hat{\epsilon}}
\providecommand{\beph}{\hat{\boldsymbol{\epsilon}}}
\providecommand{\bg}{\boldsymbol{\gamma}}
\providecommand{\bm}{\boldsymbol{\mu}}
\providecommand{\mh}{\hat{\mu}}
\providecommand{\bmh}{\hat{\boldsymbol{\mu}}}
\providecommand{\tbm}{\tilde{\boldsymbol{\mu}}}
\providecommand{\bl}{\boldsymbol{\lambda}}
\providecommand{\lh}{\hat{\lambda}}
\providecommand{\bL}{\boldsymbol{\Lambda}}
\providecommand{\tn}{\tilde{\nu}}
\providecommand{\bO}{\boldsymbol{\Omega}}
\providecommand{\bp}{\boldsymbol{\pi}}
\providecommand{\ph}{\hat{\pi}}
\providecommand{\bph}{\hat{\boldsymbol{\pi}}}
\providecommand{\sh}{\hat{\sigma}}
\providecommand{\bS}{\boldsymbol{\Sigma}}
\providecommand{\bSh}{\hat{\boldsymbol{\Sigma}}}
\providecommand{\bt}{\boldsymbol{\theta}}
\renewcommand{\th}{\hat{\theta}}
\providecommand{\bth}{\hat{\boldsymbol{\theta}}}
\providecommand{\tth}{\tilde{\theta}}
\providecommand{\bta}{\boldsymbol{\tau}}
\providecommand{\tah}{\hat{\tau}}
\providecommand{\btah}{\hat{\boldsymbol{\tau}}}

%% Subscripts/tildes
\providecommand{\tb}{\widetilde{\beta}}
\providecommand{\tbb}{\widetilde{\boldsymbol{\beta}}}
\providecommand{\bbj}{\boldsymbol{\beta}_{-j}}
\providecommand{\bbhj}{\widehat{\boldsymbol{\beta}}_{-j}}
\providecommand{\rj}{\mathbf{r}_{-j}}
\providecommand{\rr}{\tilde{\mathbf{r}}}
\providecommand{\xj}{\mathbf{x}_{-j}}
\providecommand{\Xj}{\mathbf{X}_{-j}}
\providecommand{\XX}{\widetilde{\mathbf{X}}}
\providecommand{\WW}{\widetilde{\mathbf{W}}}
\providecommand{\yy}{\tilde{\mathbf{y}}}
\providecommand{\yh}{\hat{y}}
\providecommand{\byh}{\hat{\mathbf{y}}}
\providecommand{\zz}{\tilde{\mathbf{z}}}
\providecommand{\ZZ}{\tilde{\mathbf{Z}}}

%% Operators
\renewcommand{\Pr}{\mathbb{P}}
\providecommand{\pr}{\textrm{Pr}}
\providecommand{\Ex}{\mathbb{E}}
\providecommand{\ex}{\textrm{E}}
\providecommand{\exh}{\widehat{\textrm{E}}}
\providecommand{\Var}{\mathbb{V}}
\providecommand{\var}{\textrm{Var}}
\providecommand{\varh}{\widehat{\textrm{Var}}}
\providecommand{\cov}{\textrm{Cov}}
\providecommand{\cor}{\textrm{Cor}}
\providecommand{\tr}{\textrm{tr}}
\providecommand{\Norm}{\textrm{N}}
\providecommand{\Gmma}{\textrm{Gamma}}
\providecommand{\Beta}{\textrm{Beta}}
\providecommand{\Pois}{\textrm{Pois}}
\providecommand{\NBin}{\textrm{NegBin}}
\providecommand{\Unif}{\textrm{Unif}}
\providecommand{\Binom}{\textrm{Binom}}
\providecommand{\Scx}{\textrm{Scaled-}\chi^2}
\providecommand{\Wish}{\textrm{Wishart}}
\providecommand{\Multinom}{\textrm{Multinom}}
\providecommand{\Dirch}{\textrm{Dir}}
\providecommand{\Exp}{\textrm{Exp}}

%% Statistical
\providecommand{\IID}{\overset{\text{iid}}{\sim}}
\renewcommand{\l}{\ell}
\providecommand{\inP}{\overset{\text{P}}{\longrightarrow}}
\providecommand{\inAS}{\overset{\text{a.s.\,}}{\longrightarrow}}
\providecommand{\inD}{\overset{\text{d}}{\longrightarrow}}
\providecommand{\adist}{\overset{\text{.}}{\sim}}
\providecommand{\RSS}{\textrm{RSS}}
\providecommand{\AIC}{\textrm{AIC}}
\providecommand{\BIC}{\textrm{BIC}}
\providecommand{\GCV}{\textrm{GCV}}
\providecommand{\FDR}{\textrm{FDR}}
\providecommand{\FDRh}{\widehat{\textrm{FDR}}}
\providecommand{\SNR}{\textrm{SNR}}
\providecommand{\SE}{\textrm{SE}}
\providecommand{\SD}{\textrm{SD}}
\providecommand{\OR}{\textrm{OR}}
\providecommand{\RR}{\textrm{RR}}
\providecommand{\CI}{\textrm{CI}}
\providecommand{\df}{\textrm{df}}
\providecommand{\odds}{\textrm{odds}}
\providecommand{\loglik}{\textrm{loglik}}
\providecommand{\logit}{\textrm{logit}}
\providecommand{\log}{\textrm{log}}
\providecommand{\MLE}{\textrm{MLE}}

%% Mathematical
\providecommand{\intii}{\int_{-\infty}^{\infty}}
\providecommand{\real}{\mathbb{R}}
\providecommand{\dl}{d_-}
\providecommand{\dr}{d_+}
\providecommand{\sign}{\textrm{sign}}
\providecommand{\abs}[1]{\left\lvert#1\right\rvert}
\providecommand{\norm}[1]{\left\lVert#1\right\rVert}
\providecommand{\shortnorm}[1]{\lVert#1\rVert}

\providecommand{\argmax}[1]{\underset{#1}{\operatorname{arg} \operatorname{max}}\;}
\providecommand{\argmin}[1]{\underset{#1}{\operatorname{arg} \operatorname{min}}\;}

%\newcommand{\argmin}{\operatornamewithlimits{argmin}}
%\newcommand{\argmax}{\operatornamewithlimits{argmax}}

%% Equations
\providecommand{\al}[2]{\begin{align}\label{#1}#2\end{align}}
\providecommand{\as}[1]{\begin{align*}#1\end{align*}}
\providecommand{\als}[2]{\begin{align}\label{#1}\begin{split}#2\end{split}\end{align}}
%% include your article here, just as usual
%% Note that you should use the \pkg{}, \proglang{} and \code{} commands.

\section{Introduction}
%% Note: If there is markup in \(sub)section, then it has to be escape as above.

The lasso model proposed by \cite{Tibshirani1996} has fundamentally reshaped the landscape of high-dimensional statistical research. Since its original proposal, the lasso has attracted extensive studies with a wide range of applications to many areas, such as signal processing \citep{angelosante2009rls}, gene expression data analysis \citep{huang2003linear}, face recognition \citep{wright2009robust}, text mining \citep{li2015relevance} and so on. The great success of the lasso has made it one of the most popular tools in statistical and machine-learning practice.

Recent years have seen the evolving era of Big Data where ultrahigh-dimensional, large-scale data sets are increasingly seen in many areas such as genetics, genomics, biomedical imaging, social media analysis, and high-frequency finance \citep{fan2014challenges}. Such data sets pose a challenge to solving the lasso efficiently in general, and for \proglang{R} specifically, since \proglang{R} is not naturally well-suited for analyzing large-scale data sets \citep{kane2013scalable}. Thus, there is a clear need for scalable software for fitting lasso-type models designed to meet the needs of big data.

In this project, we develop an \proglang{R} package, \pkg{biglasso} \citep{biglasso}, to extend lasso model fitting to Big Data in \proglang{R}. Specifically, sparse linear and logistic regression models with lasso and elastic net penalties are implemented. The most notable features of \pkg{biglasso} include:
\begin{itemize}
\item It utilizes memory-mapped files to store the massive data on the disk, only loading data into memory when necessary during model fitting. Consequently, it's able to seamlessly handle out-of-core computation.
\item It is built upon pathwise coordinate descent algorithm and ``warm start'' strategy, which has been proven to be one of fastest approaches to solving the lasso \citep{friedman2010regularization}.
\item We develop new, hybrid feature screening rules that outperform state-of-the-art screening rules such as the sequential strong rule (SSR) \citep{tibshirani2012strong}, and the sequential EDPP rule (SEDPP) \citep{JMLR:v16:wang15a} with additional 1.5x to 4x speedup.
\item The implementation is designed to be as memory-efficient as possible by eliminating extra copies of the data created by other \proglang{R} packages, making \pkg{biglasso} at least 2x more memory-efficient than \pkg{glmnet}.
\item The underlying computation is implemented in \proglang{C++}, and parallel computing with \proglang{OpenMP} is also supported.
\end{itemize}

The methodological innovation and well-designed implementation have made \pkg{biglasso} a much more memory- and computation-efficient and highly scalable lasso solver, as compared to existing popular \proglang{R} packages like \pkg{glmnet} \citep{friedman2010regularization}, \pkg{ncvreg} \citep{Breheny2011}, and \pkg{picasso} \citep{picasso}. More importantly, to the best of our knowledge, \pkg{biglasso} is the first \proglang{R} package that enables the user to fit lasso models with data sets that are larger than available RAM, thus allowing for powerful big data analysis on an ordinary laptop.

The rest of the paper is organized as follows. In Section \ref{sect:method}, we describe the memory-mapping technique for out-of-core computation as well as our newly developed hybrid safe-strong rule for feature screening. In Section \ref{sect:implement}, we discuss some important implementation techniques as well as parallel computation that make \pkg{biglasso} memory- and computation-efficient. Section \ref{sect:experiment} presents benchmarking experiments with both simulated and real data sets. Section \ref{sect:example} provides a brief, reproducible demonstration illustrating how to use \pkg{biglasso}, while \ref{sect:application} demonstrates the out-of-core computation capability of \pkg{biglasso} through its application to a large-scale genome-wide association study. We conclude the paper with some final discussions in Section \ref{sect:conclusion}.

\section{Method} \label{sect:method}

%\subsubsection{Notation}
%
%We denote $\X = (\x_1, \ldots, \x_p)$ be the $n \times p$ feature matrix with cell $x_{ij}$,  $\y \in \mathbb{R}^n$ the response vector, $\r \in \mathbb{R}^n$ the residual vector, and $\langle \cdot, \cdot \rangle$ the inner-product of two vectors. The standardized feature matrix is denoted as $\tilde{\X}$ such that $\sum_{i=1}^n \tilde{x}_{ij} = 0, \; \frac{1}{n}\sum_{i=1}^n \tilde{x}_{ij}^2 = 1, \; j = 1, \ldots, p.$

\subsection{Memory mapping}

\textit{Memory mapping} \citep{bovet2005understanding} is a technique that maps a data file into the virtual memory space so that the data on the disk can be accessed as if they were in the main memory. Technically, when the program starts, the operating system (OS) will cache the data into RAM. Once the data are in RAM, the computation is at the standard in-memory speed. If the program requests more data after the memory is fully occupied, which is inevitable in the data-larger-than-RAM case, the OS will move data that is not currently needed out of RAM to create space for loading in new data. This is called the \textit{page-in-page-out} procedure, and is automatically handled by the OS.

The memory mapping technique is commonly used in modern operating systems such as Windows and Unix-like systems due to several advantages:
%\citep{wiki:mmap}.
%From the computational perspective, memory mapping is particularly useful for big data computation because: 
\begin{enumerate}[(1)]
\item it provides faster file read/write than traditional I/O methods since data-copy from kernel to user buffer is not needed due to page caches;
\item it allows random access to the data as if it were in the main memory even though it physically resides on the disk;
\item it supports concurrent sharing in that multiple processes can access the same memory-mapped data file simultaneously, making parallel computing easy to implement in data-larger-than-RAM cases;
\item it enables out-of-core computing thanks to the automatic page-in-page-out procedure.
\end{enumerate}
%\begin{enumerate}[(1)]
%\item it provides faster file read/write than traditional I/O methods since data-copy from kernel to user buffer is not needed due to page caches;
%\item it allows random access to the data file as if the data were in the main memory even it's on the disk physically;
%\item it supports concurrent sharing in that multiple processes can access the same memory-mapped data file simultaneously, making parallel computing easy to implement in data-larger-than-RAM cases;
%\item it enables out-of-core computing thanks to the automatic ``page-in-page-out'' procedure.
%\end{enumerate}
We refer the readers to \cite{rao2010critical}, \cite{lin2014mmap}, and \cite{bovet2005understanding} for detailed techniques and some successful applications of memory mapping.
%We refer the readers to \cite{lin2014mmap} and \cite{bovet2005understanding} for details of memory mapping.

To take advantage of memory mapping, \pkg{biglasso} creates memory-mapped big matrix objects based upon the R package \pkg{bigmemory} \citep{kane2013scalable}, which uses the Boost C++ library and implements memory-mapped big matrix objects that can be directly used in \proglang{R}. Then at the \proglang{C++} level, \pkg{biglasso} uses the \proglang{C++} library of \pkg{bigmemory} for underlying computation and model fitting.

\subsection{Efficient feature screening}

Another important contribution of \pkg{biglasso} is our newly developed \textit{hybrid safe-strong rule}, named SSR-BEDPP, which substantially outperforms existing state-of-the-art ones in terms of the overall computing time of obtaining the lasso solution path.  Here, we describe the main idea of hybrid rules; for the technical details, see \citet{ZengRules}.

\textit{Feature screening} aims to identify and discard inactive features (i.e., those with zero coefficients) from the lasso optimization. It often leads to dramatic dimension reduction and hence significant computation savings. However, these savings will be negated if the screening rule itself is too complicated to execute. Therefore, an efficient screening rule needs to be powerful enough to discard a large portion of features and also relatively simple to compute.

\begin{figure}[h]
\centering
\includegraphics[scale=0.4]{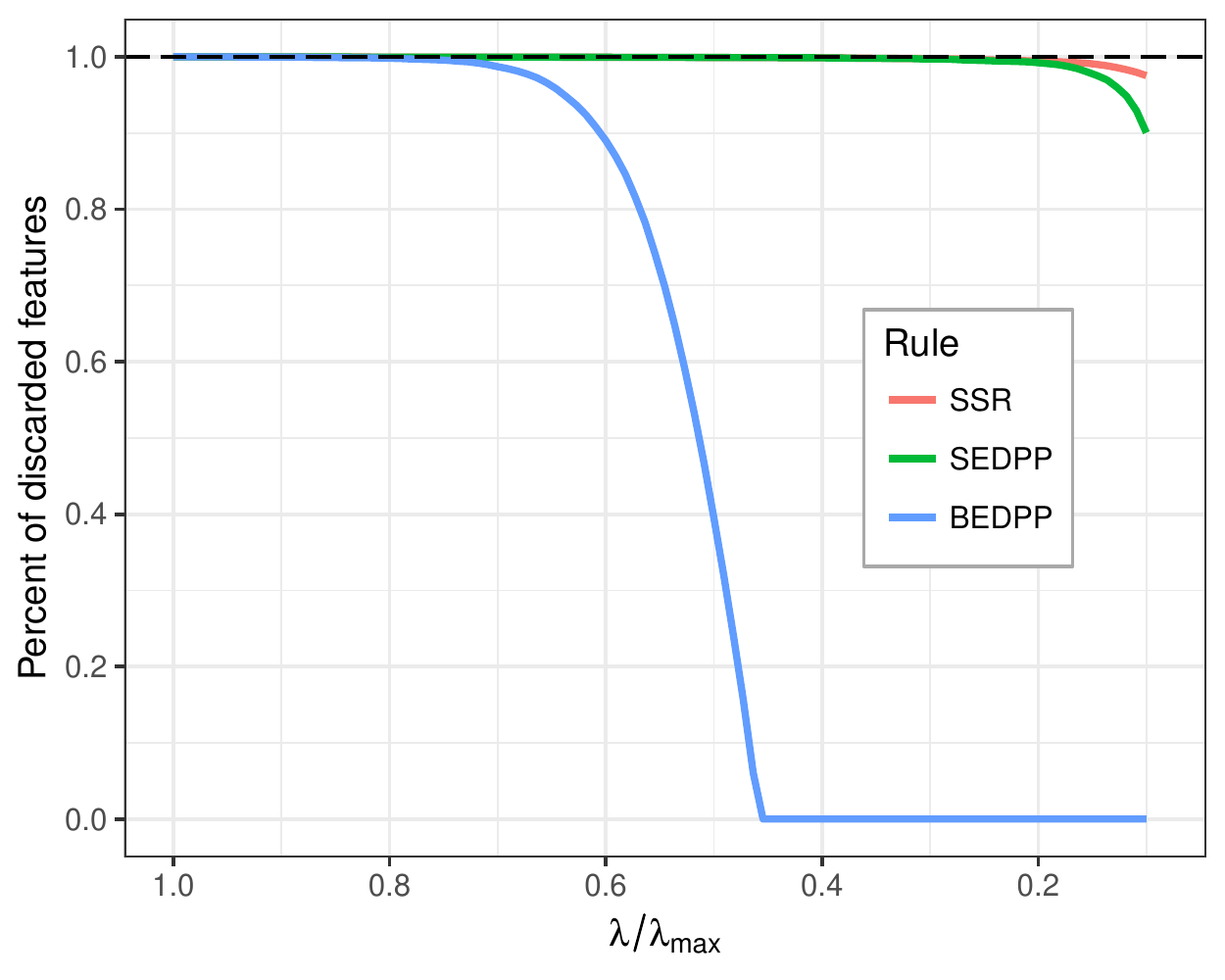}
\caption{Percent of features discarded.}
\label{fig:rule_compare}
\end{figure}

Existing screening rules for the lasso can be divided into two types: (1) heuristic rules, such as the sequential strong rule (SSR) \citep{tibshirani2012strong}, and (2) safe rules, such as the basic and the sequential EDPP rules \citep{JMLR:v16:wang15a}, denoted here as BEDPP and SEDPP respectively. Safe rules, unlike heuristic ones, are guaranteed to never incorrectly screen a feature with a nonzero coefficient.  Figure \ref{fig:rule_compare} compares the power of the three rules in discarding features. SSR, though most powerful among the three, requires a cumbersome post-convergence check to verify that it has not incorrectly discarded an active feature.  The SEDPP rule is both safe and powerful, but is inherently complicated and time-consuming to evaluate. Finally, BEDPP is the least powerful, and discards virtually no features when $\lambda$ is smaller than 0.45 (in this case), but is both safe and involves minimal computational burden.

The rule employed by \pkg{biglasso}, SSR-BEDPP, as its name indicates, combines
SSR with the simple yet safe BEDPP rule. The rationale is to alleviate the
burden of post-convergence checking for strong rules by not checking features
that can be safely eliminated using BEDPP.  This hybrid approach leverages the
advantages of each rule, and offers substantial gains in efficiency,
especially when solving the lasso for large values of $\lambda$.

%\begin{table}[h]
%\caption{Complexity of computing screening rules along the entire path of $\mathcal{K}$ values of $\lambda$. $| \mathcal{S}_k |$ denotes the cardinality of safe set $\mathcal{S}_k$; i.e., the set of features not discarded by BEDPP screening.}
%\centering
%\begin{threeparttable}
%\begin{adjustbox}{max width=\linewidth, keepaspectratio, center}
%\begin{tabular}{rcc}
%\hline
%Rule & Operations & Complexity\\
%\hline
%SSR &  $np \mathcal{K}$\tnote{*} & $O(np \mathcal{K})$\\
%SEDPP & $3np \mathcal{K}$ & $O(np \mathcal{K})$ \\
%BEDPP & $p\mathcal{K} + 2np$ & $O(np)$\\
%SSR-BEDPP & $n \sum_k^\mathcal{K} |\mathcal{S}_k| + p\mathcal{K} + 2np$ & $O(np + n\sum_k^\mathcal{K} |\mathcal{S}_k|))$\\
%\hline
%\end{tabular}
%\end{adjustbox}
%\begin{tablenotes}\footnotesize
%\item[*] Post-convergence checking included. Assumes no violations.
%\end{tablenotes}
%\end{threeparttable}
%\label{tab_complexity}
%\end{table}

Table \ref{tab_complexity} summarizes the complexities of the four rules when applied to solving the lasso along a path of $K$ values of $\lambda$ for a data set with $n$ instances and $p$ features.  SSR-BEDPP can be substantially faster than the other three rules when BEDPP is effective. Furthermore, it is important to note that SSR (with post-convergence checking) and SEDPP have to scan the entire feature matrix at every value of $\lam$, while SSR-BEDPP only needs to scan the features not discarded by BEDPP.  This advantage of SSR-BEDPP is particularly appealing in out-of-core computing, where fully scanning the feature matrix requires disk access and therefore becomes the computational bottleneck of the procedure.
\begin{table}[h]
\centering
%\begin{adjustbox}{max width=\linewidth, keepaspectratio, center}
\begin{tabular}{rc}
\toprule
Rule & Complexity\\
\midrule
SSR &  $O(np K)$\\
SEDPP & $O(np K)$ \\
BEDPP & $O(np)$\\
%SSR-BEDPP & $O(np + n\sum_k^K |\mathcal{S}_k|))$\\
SSR-BEDPP & $O(n\sum_k^K |\mathcal{S}_k|))$\\
\bottomrule
\end{tabular}
%\end{adjustbox}
\caption{Complexity of computing screening rules along the entire path of $K$ values of $\lambda$. $| \mathcal{S}_k |$ denotes the cardinality of $\mathcal{S}_k$, the safe set of features not discarded by BEDPP screening.}
\label{tab_complexity}
\end{table}
%\subsection{Pathwise coordinate optimization}
%
%Like packages \pkg{glmnet} and \pkg{ncvreg}, our package \pkg{biglasso} implements the pathwise coordinate descent algorithm \citep{friedman2007pathwise} to solve the lasso along an entire path of values of the regularization parameter $\lambda$. This solution-path algorithm makes use of previous solution $\widehat{\bb}(\lambda_{k-1})$ as warm starts for solving $\widehat{\bb}(\lambda_k)$, making the algorithm very efficient. Moreover, ``active set cycling'' and feature screening strategies are also incorporated to boost the optimization by iterating over a subset of features. Due to space limitatio, we refer the readers to \cite{friedman2010regularization} for the algorithm details.

The hybrid screening idea is straightforward to extend to other lasso-type problems provided that a corresponding safe rule exists. For the \pkg{biglasso} package, we also implemented a hybrid screening rule, SSR-Slores, for lasso-penalized logistic regression by combining SSR with the so-called Slores rule~\citep{wang2014safe}, a safe screening rule developed for sparse logistic regression.

\section{Implementation} \label{sect:implement}

\subsection{Memory-efficient design}

In penalized regression models, the feature matrix $\X \in \mathbb{R}^{n \times p}$ is typically standardized to ensure that the penalty is applied uniformly across features with different scales of measurement. In addition, standardization contributes to faster convergence of the optimization algorithm.  In existing \proglang{R} packages such as \pkg{glmnet}, \pkg{ncvreg}, and \pkg{picasso}, a standardized feature matrix $\widetilde{\X}$ is calculated and stored, effectively doubling memory usage. This problem is compounded by cross-validation, where these packages also calculate and store additional standardized and unstandardized copies of $\X$ for each fold. This approach does not scale up well for big data.

To make the memory usage more efficient, \pkg{biglasso} doesn't store $\widetilde{\X}$. Instead, it saves only the means and standard deviations of the columns of $\X$ as two vectors, denoted as $\c$ and $\s$. Then wherever $\widetilde{x}_{ij}$ is needed, it is retrieved by ``cell-wise standardization'', i.e., $\widetilde{x}_{ij} = (x_{ij} - c_j) / s_j$. Additionally, the estimated coefficient matrix is sparse-coded in \proglang{C++} and \proglang{R} to save memory space.

\subsubsection{Simplification of computations}

Cell-wise standardization saves a great deal of memory space, but at the expense of computational efficiency. To minimize this, \pkg{biglasso} uses a number of computational strategies to eliminate redundant calculations.

We first note that the computations related to $\widetilde{\X}$ during whole model fitting process are mainly of three types, and all can be simplified so that na\"ive cell-wise standardization can be avoided: 
\begin{enumerate}[(1)]
\item $\widetilde{\x}_j^T \widetilde{\x}_* = \sum_i \frac{x_{ij} - c_j}{s_j} \frac{x_{i*} -c_*}{s_*} = \frac{1}{s_j s_*} \big( \sum_i x_{ij} x_{i*} - n c_j c_* \big)$;
\item $\widetilde{\x}_i^T \y = \sum_i \frac{x_{ij} - c_j}{s_j} y_i  = \frac{1}{s_j} \big( \sum_i x_{ij} y_i - c_j  \sum_i y_i\big)$;
\item $\widetilde{\x}_j^T \r = \sum_i \frac{x_{ij} - c_j}{s_j} r_i = \frac{1}{s_j} \big(\sum_i x_{ij} r_i - c_j \sum_i r_i \big)$;
%\item $\widetilde{\x}_j^T \W \r = \frac{1}{s_j} \left( \sum_i w_{ii} x_{ij} r_i - c_j \sum_i w_{ii} r_i \right)$; 
%\item $\widetilde{\x}_j^T \W \widetilde{\x}_j = \sum_i w_{ii} \left( \frac{x_{ij} - c_j}{s_j} \right)^2 = \frac{1}{s_j^2} \left \{ \sum_i w_{ii} x_{ij}^2 -2 c_j \sum_i w_{ii} x_{ij} + c_j^2 \sum_i w_{ii} \right \}$
%\item $\widetilde{\x}_j^T \W \widetilde{\x}_j =  \frac{1}{s_j^2} \left( \sum_i w_{ii} x_{ij}^2 -2 c_j \sum_i w_{ii} x_{ij} + c_j^2 \sum_i w_{ii} \right)$.
\end{enumerate}
where $\widetilde{\x}_j$ is the $j$th column of $\widetilde{\X}$, $\widetilde{\x}_*$ is the column corresponding to $\lambda_{\max}$, $\y$ is the response vector, and $\r \in \mathbb{R}^n$ is current residual vector.

Type (1) and (2) are used only for initial feature screening, and require only one-time execution. Type (3) occurs in both the coordinate descent algorithm and the post-convergence checking. Since the coordinate descent algorithm is fast to converge and only iterates over features in the active set $\mathcal{A}$ of nonzero coefficients, whose size is much smaller than $p$, the number of additional computations this introduces is small. Moreover, we pre-compute and store $\sum_i r_i$, which saves a great deal of computation during post-convergence checking since $\r$ does not change during this step. As a result, our implementation of cell-wise standardization requires only $O(p)$ additional operations compared to storing the entire standardized matrix.

%[YZ: 
%
%I made some changes here:
%0) I added type (2): X^T y
%1) Previously type (1), (3) calculations were not correct, need parentheses;
%2) I changed Type (1) to be X^T x_*, matrix-vector multiplication. Previous type (1) would look like matrix-matrix multiplication. 
%3) Also, the additional operations should be O(p) instead of O(1) if considering the computations for all columns.

%Feel free to make changes and let me know if there are any questions.
%]

\subsubsection{Scalable cross-validation}

%The cross-validation is proceeded by randomly partitioning the data into training and test sets, fitting a model with training set, and then assessing the fitted model on the test set with respect to a performance metric. This procedure is usually done multiple times. 

Cross-validation is integral to lasso modeling in practice, as it is by far the most common approach to choosing $\lam$. It requires splitting the data matrix $\X$ into training and test sub matrices, and fitting the lasso model multiple times. This procedure is also memory-intensive, especially if performed in parallel.

Existing lasso-fitting \proglang{R} packages split $\X$ using the ``slicing operator'' directly in \proglang{R} (e.g., \texttt{X[1:1000, ]}). This introduces a great deal of overhead and hence is quite slow when $\X$ is large. Worse, the training and test sub-matrices must be saved into memory, as well as their standardized versions, all of which result in considerable memory consumption. 

In contrast, \pkg{biglasso} implements a much more memory-efficient cross-validation procedure that avoids the above issues. The key design is that the main model-fitting \proglang{R} function allows a subset of $\X$, indicated by the row indices, as input. To cope with this design, all underlying \proglang{C++} functions are enabled to operate on a subset of $\X$ given a row-index vector is provided.

Consequently, instead of creating and storing sub-matrices, only the indices of the training/test sets and the descriptor of $\X$ (essentially, an external pointer to $\X$) are needed for parallel cross validation thanks to the concurrency of memory-mapping. The net effect is that only one memory-mapped data matrix $\X$ is needed for $K$-fold parallel cross-validation, whereas other packages need (roughly) $2K$ copies of $\X$, a copy and a standardized copy for each fold.

\subsection{Parallel computation}

Another important feature of \pkg{biglasso} is its parallel computation capability. There are two types of parallel computation implemented in \pkg{biglasso}.

At the \proglang{C++} level, single model fitting (as opposed to cross validation) is parallelized with OpenMP. Though the pathwise coordinate descent algorithm is inherently sequential and thus does not lend itself to parallelization, several components of the algorithm (computing $\c$ and $\s$, matrix-vector multiplication, post-convergence checking, feature screening, etc.) do, and are parallel-enabled in \pkg{biglasso}.

%% At the R level, the cross-validation procedure is naturally parallelized using the R package \pkg{parallel}, as in \pkg{glmnet} and \pkg{ncvreg}. The advantage of \pkg{biglasso} is that it is much more memory- and computation-efficient by avoiding extra data copies and heavy overhead associated with copying data to parallel workers. Note that when cross-validation is run in parallel in R, parallel computing at C++ level for single model-fitting is disabled to avoid nested parallelization.

Parallelization can also be implemented at the \proglang{R} level to run cross-validation in parallel.  This implementation is straightforward and also implemented by \pkg{ncvreg} and \pkg{glmnet}.  However, as mentioned earlier, the parallel implementation of \pkg{biglasso} is much more memory- and computation-efficient by avoiding extra copies and the overhead associated with copying data to parallel workers. Note that when cross-validation is run in parallel in \proglang{R}, parallel computing at \proglang{C++} level for single model-fitting is disabled to avoid nested parallelization.

%\section{Logistic regression} \label{sect:log_reg}
%
%\begin{itemize}
%\item no hybrid safe-strong rule so far;
%\item newton-CD and Majorization-minimization-CD
%\end{itemize}

%\section{Numerical results} \label{sect:experiment}
\section{Benchmarking experiments} \label{sect:experiment}

In this section, we demonstrate that our package \pkg{biglasso} (1.2-3) is considerably more efficient at solving for lasso estimates than existing popular \proglang{R} packages \pkg{glmnet} (2.0-5), \pkg{ncvreg} (3.9-0), and \pkg{picasso} (0.5-4). Here we focus on solving lasso-penalized linear and logistic regression, respectively, over the entire path of 100 $\lambda$ values which are equally spaced on the scale of $\lambda / \lambda_{\max}$ from 0.1 to 1.  To ensure a fair comparison, we set the convergence thresholds to be equivalent across all four packages.  All experiments are conducted with 20 replications, and the average computing times (in seconds) are reported. The benchmarking platform is a MacBook Pro with Intel Core i7 @ 2.3 GHz and 16 GB RAM. 

\subsection{Memory efficiency}

To demonstrate the improved memory efficiency of \pkg{biglasso} compared to existing packages, we simulate a feature matrix with dimensions $1,000 \times 100,000$. The raw data is 0.75 GB, and stored on the hard drive as an \proglang{R} data file and a memory-mapped file. We used \texttt{Syrupy}\footnote{\url{https://github.com/jeetsukumaran/Syrupy}} to measure the memory used in RAM (i.e. the resident set size, RSS) every 1 second during lasso-penalized linear regression model fitting by each of the packages. 

The maximum RSS during the model fitting is reported in Table \ref{tab_memo}. In the single fit case, \pkg{biglasso} consumes 0.84 GB memory in RAM, 50\% of that used by \pkg{glmnet} and  22\% of that used by \pkg{picasso}. Note that the memory consumed by \pkg{glmnet}, \pkg{ncvreg}, and \pkg{picasso} are respectively 2.2x,  2.1x, and 5.1x larger than the size of the raw data.

More strikingly, \pkg{biglasso} does not require additional memory to perform cross-validation, unlike other packages.  For serial 10-fold cross-validation, \pkg{biglasso}  requires just 27\% of the memory used by \pkg{glmnet} and 23\% of that used by \pkg{ncvreg}, making it 3.6x and 4.3x more memory-efficient than \pkg{glmnet} and \pkg{ncvreg}, respectively.

The memory savings offered by \pkg{biglasso} would be even more significant if cross-validation were conducted in parallel. However, measuring memory usage across parallel processes is not straightforward and not implemented in \texttt{Syrupy}.

%\begin{table}[h] 
%\caption{The maximum RSS (in MB) by the four packages for a single fit and 10 fold cross-validation (CV).}
%\centering
%\begin{threeparttable}
%\begin{adjustbox}{max width=\linewidth, center}
%\begin{tabular}{r|cccc}
%\toprule
%%\hline
%Package & picasso\tnote{*} & ncvreg & glmnet & biglasso \\
%\midrule
%Single fit & 3933 & 1640 & 1707 & 865 \\
%%4-fold CV (1 core) &  - &  2762 & 3194  & 893 \\
%10-fold CV (1 core) &  - &  3827 & 3261  & 894 \\
%%10-fold CV (4 cores) &  - &   &   &  \\
%\bottomrule
%\end{tabular}
%\end{adjustbox}
%\begin{tablenotes}\footnotesize
%\item[*] Cross-validation is not implemented in \pkg{picasso}.
%\end{tablenotes}
%\end{threeparttable}
%\label{tab_memo}
%\end{table}
\begin{table}[h] 
\centering
\begin{threeparttable}
%\begin{adjustbox}{max width=\linewidth, center}
\begin{tabular}{r|cccc}
\toprule
%\hline
Package & picasso\tnote{*} & ncvreg & glmnet & biglasso \\
\midrule
Single fit & 3.84 & 1.60 & 1.67 & 0.84 \\
%4-fold CV (1 core) &  - &  2.70 & 3.12  & 0.87 \\
10-fold CV (1 core) &  - &  3.74 & 3.18  & 0.87 \\
%10-fold CV (4 cores) &  - &   &   &  \\
\bottomrule
\end{tabular}
%\end{adjustbox}
\begin{tablenotes}\footnotesize
\item[*] Cross-validation is not implemented in \pkg{picasso}.
\end{tablenotes}
\end{threeparttable}
\caption{The maximum RSS (in GB) for a single fit and 10 fold cross-validation (CV) with the raw data of 0.75 GB.}
\label{tab_memo}
\end{table}

\subsection{Computational efficiency: Linear regression}

\subsubsection{Simulated data}

We now show with simulated data that \pkg{biglasso} is more scalable in both $n$ and $p$ (i.e., number of instances and features). We adopt the same model in \cite{JMLR:v16:wang15a} to simulate data: $\y = \X \bb + 0.1 \bep$, where $\X$ and $\bep$ are i.i.d. sampled from $N(0, 1)$. We consider two different cases: (1) Case 1: varying $p$. We set $n=1,000$ and vary $p$ from 1,000 to 20,000. We randomly select 20 true features, and sample their coefficients from Unif[-1, 1]. After simulating $\X$ and $\bb$, we then generate $\y$ according to the true model; (2) Case 2: varying $n$. We set $p=10,000$ and vary $n$ from 200 to 20,000. $\bb$ and $\y$ are generated in the same way as in Case 1.

\begin{figure}[ht]
\centering
\begin{subfigure}{.5\textwidth}
  \centering
  \includegraphics[width=\linewidth]{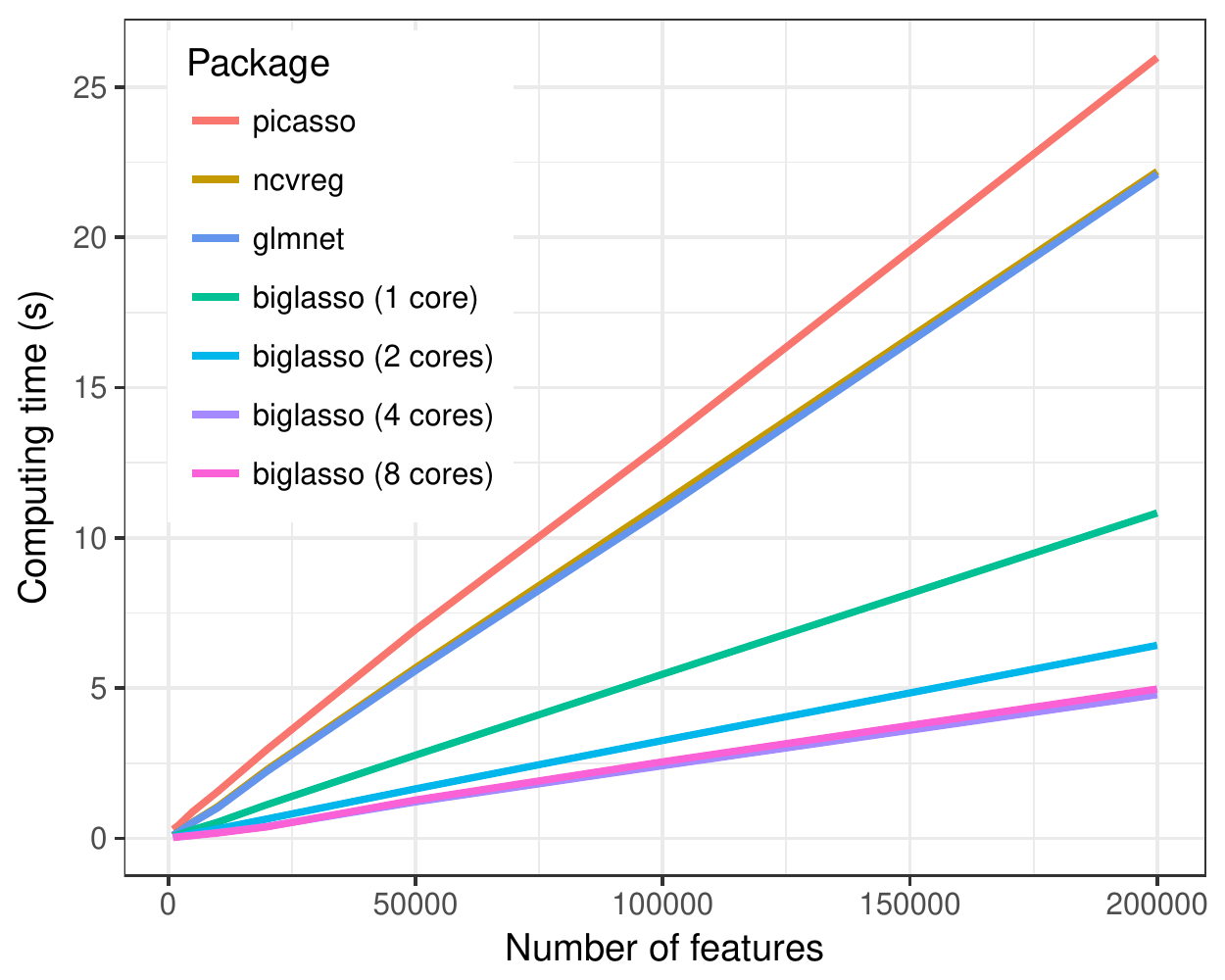}
  \caption{Varying $p$, $n = 1,000$.}
\end{subfigure}%
\begin{subfigure}{.5\textwidth}
  \centering
  \includegraphics[width=\linewidth]{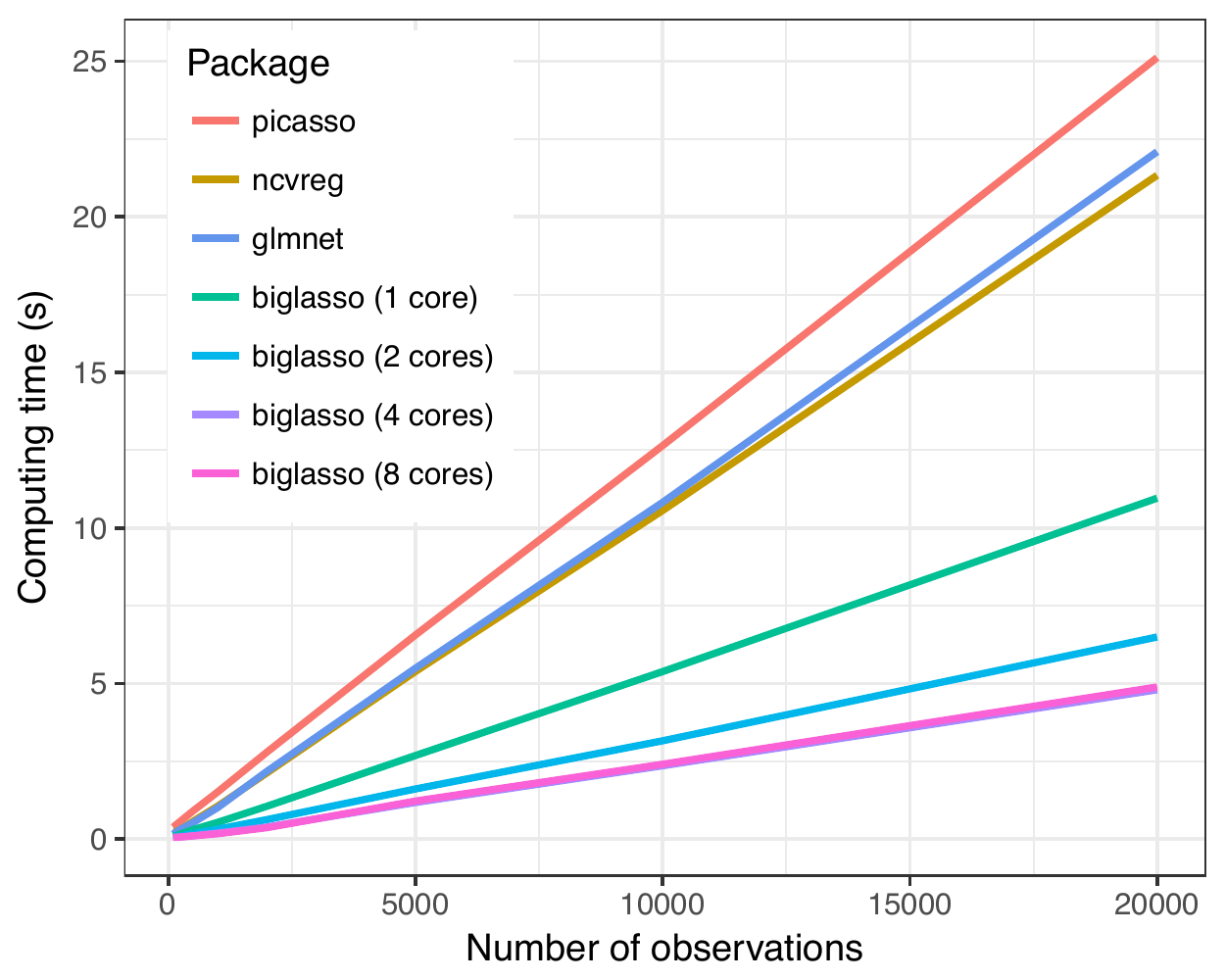}
  \caption{Varying $n$, $p = 10,000$.}
\end{subfigure}
\caption{Mean computing time (in seconds) of solving the lasso over a sequence 100 $\lambda$ values as a function of $p$ (Left) and $n$ (Right).}
\label{fig:simu_res}
\end{figure}

Figure \ref{fig:simu_res} compares the mean computing time of solving the lasso over a sequence of 100 $\lambda$ values by the four packages. In all the settings, \pkg{biglasso} (1 core) is uniformly 2x faster than \pkg{glmnet} and \pkg{ncvreg}, and 2.5x faster than \pkg{picasso}. Moreover, the computing time of \pkg{biglasso} can be further reduced by half via parallel-computation of 4 cores. Using 8 cores doesn't help due to the increased overhead of communication between cores.

%\begin{figure}[h]
%\centering
%\includegraphics[scale=0.4]{2016-11-20_vary_p_pkgs.png}
%\caption{Mean computing time (seconds) of solving the lasso over a sequence 100 $\lambda$ values as a function of $p$.}
%\label{fig_vary_p}
%\end{figure}
%
%\begin{figure}[h]
%\centering
%\includegraphics[scale=0.4]{2016-11-20_vary_n_pkgs.png}
%\caption{Mean computing time (seconds) of solving the lasso over a sequence 100 $\lambda$ values as a function of $n$.}
%\label{fig_vary_n}
%\end{figure}

\subsubsection{Real data}

In this section, we compare the performance of the packages using diverse real data sets: (1) Breast cancer gene expression data\footnote{\url{http://myweb.uiowa.edu/pbreheny/data/bcTCGA.html}} (GENE); (2) MNIST handwritten image data\footnote{\url{http://yann.lecun.com/exdb/mnist/}} (MNIST); (3) Cardiac fibrosis genome-wide association study data\footnote{\url{https://arxiv.org/abs/1607.05636}} (GWAS); and (4) Subset of New York Times bag-of-words data\footnote{\url{https://archive.ics.uci.edu/ml/datasets/Bag+of+Words}} (NYT). Note that for data sets MNIST and NYT, a different response vector is randomly sampled from a test set at each replication.

The size of the feature matrices and the average computing times are summarized in Table \ref{tab_real_res}. In all four settings, \pkg{biglasso} was fastest at obtaining solutions, providing 2x to 3.8x speedup compared to \pkg{glmnet} and \pkg{ncvreg}, and 2x to 4.6x speedup compared to \pkg{picasso}.

% old results
\begin{table}[h] 
\centering
%\begin{adjustbox}{max width=\linewidth, center}
\begin{tabular}{r|cccc}
\toprule
%\hline
Package & GENE & MNIST & GWAS & NYT \\
 & $n=536$ & $n=784$ & $n=313$ & $n=5,000$ \\
 & $p=17,322$ & $p=60,000$ & $p=660,495$ & $p=55,000$ \\
\midrule
picasso & 1.50 (0.01) & 6.86 (0.06) & 34.00 (0.47) & 44.24 (0.46) \\
ncvreg & 1.14 (0.02) & 5.60 (0.06) & 31.55 (0.18) & 32.78 (0.10) \\
glmnet & 1.02 (0.01) & 5.63 (0.05) & 23.23 (0.19) & 33.38 (0.08) \\
biglasso & 0.54 (0.01) & 1.48 (0.10) & 17.17 (0.11) & 14.35 (1.29) \\
\bottomrule
\end{tabular}
%\end{adjustbox}
\caption{Mean (SE) computing time (seconds) for solving the lasso along a sequence of 100 $\lambda$ values.}
\label{tab_real_res}
\end{table}

%% new results including parallel computing
%\begin{table}[h] 
%\centering
%%\begin{adjustbox}{max width=\linewidth, center}
%\begin{tabular}{r|cccc}
%\toprule
%%\hline
%Package & GENE & MNIST & GWAS & NYT \\
% & $n=536$ & $n=784$ & $n=313$ & $n=5,000$ \\
% & $p=17,322$ & $p=60,000$ & $p=660,495$ & $p=55,000$ \\
%\midrule
%picasso & 1.77 (0.02) & 6.90 (0.03) & 36.37 (0.55) & 44.64 (0.45) \\
%ncvreg & 1.21 (0.01) & 5.84 (0.03) & 32.43 (0.15) & 35.79 (0.08) \\
%glmnet & 1.16 (0.01) & 5.70 (0.04) & 24.43 (0.17) & 33.51 (0.07) \\
%biglasso (1 core) & 0.68 (0.01) & 1.53 (0.10) & 18.19 (0.02) & 19.30 (0.98) \\
%biglasso (2 cores) & 0.44 (0.01) & 1.84 (0.07) & 14.70 (0.02) & 17.21 (0.61)\\
%biglasso (4 cores) & 0.34 (0.01) & 1.42 (0.05) & 11.80 (0.04) & 13.88 (0.58)\\
%biglasso (8 cores) & 0.37 (0.01) & 1.53 (0.05) & 12.06 (0.03) & 14.64 (0.66)\\
%\bottomrule
%\end{tabular}
%%\end{adjustbox}
%\caption{Mean (SE) computing time (seconds) for solving the lasso along a sequence of 100 $\lambda$ values.}
%\label{tab_real_res}
%\end{table}

\subsection{Computational efficiency: Logistic regression}

\subsubsection{Simulated data}

Similar to Section 4.2, here we first illustrate that \pkg{biglasso} is faster than other packages in fitting the logistic regression model with simulated data. The true data-generating model is: $y_i \sim Bin(1, prob); \logit(prob) = \x_i \bb$, where each entry of $\x_i$ is i.i.d. sampled from standard Gaussian distribution. Again, two cases -- varying $p$ and varying $n$ -- are considered. 20 true features are randomly chosen and their coefficients are sampled from Unif[-1, 1].

\begin{figure}[ht]
\centering
\begin{subfigure}{.5\textwidth}
  \centering
  \includegraphics[width=\linewidth]{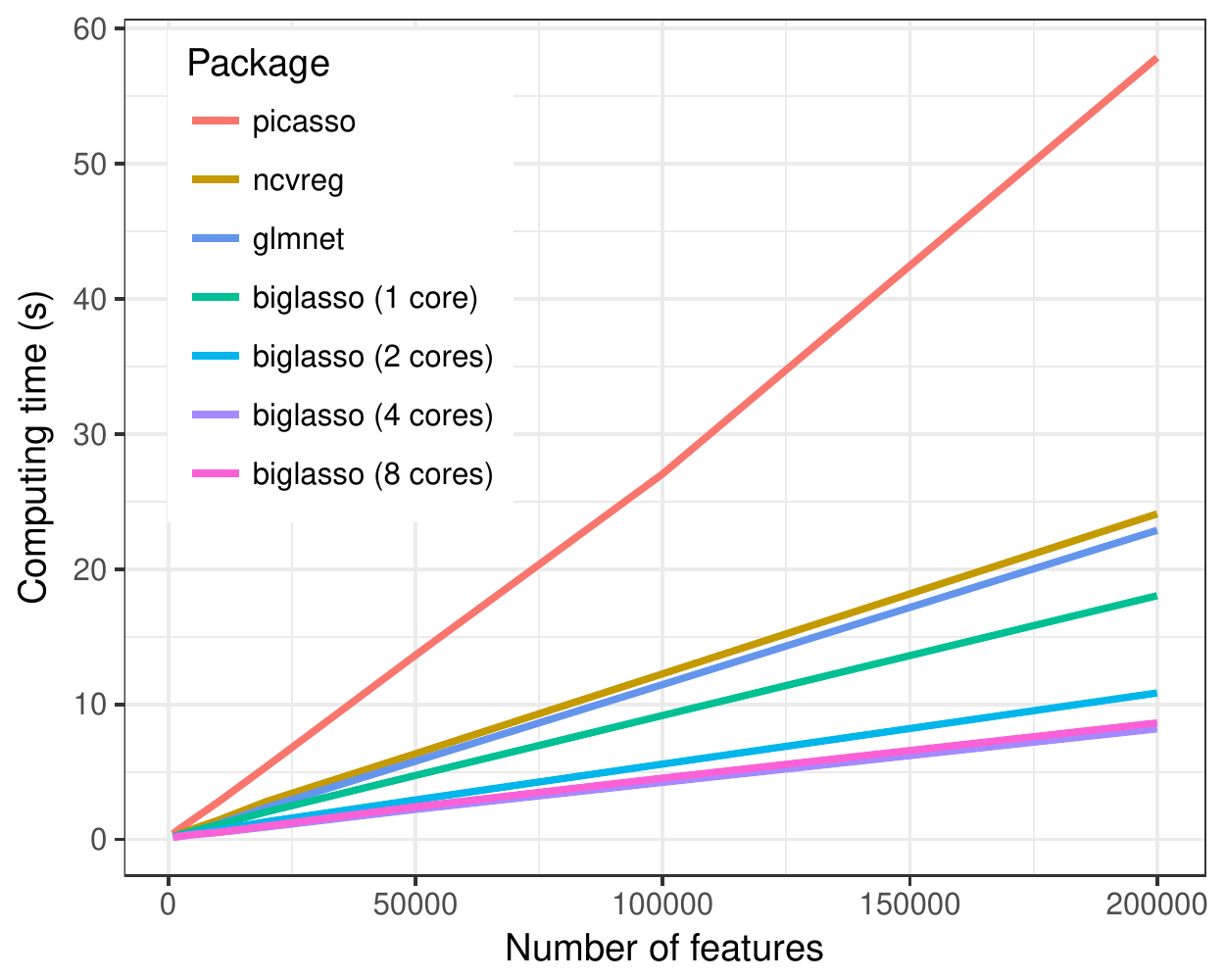}
  \caption{Varying $p$, $n = 1,000$.}
\end{subfigure}%
\begin{subfigure}{.5\textwidth}
  \centering
  \includegraphics[width=\linewidth]{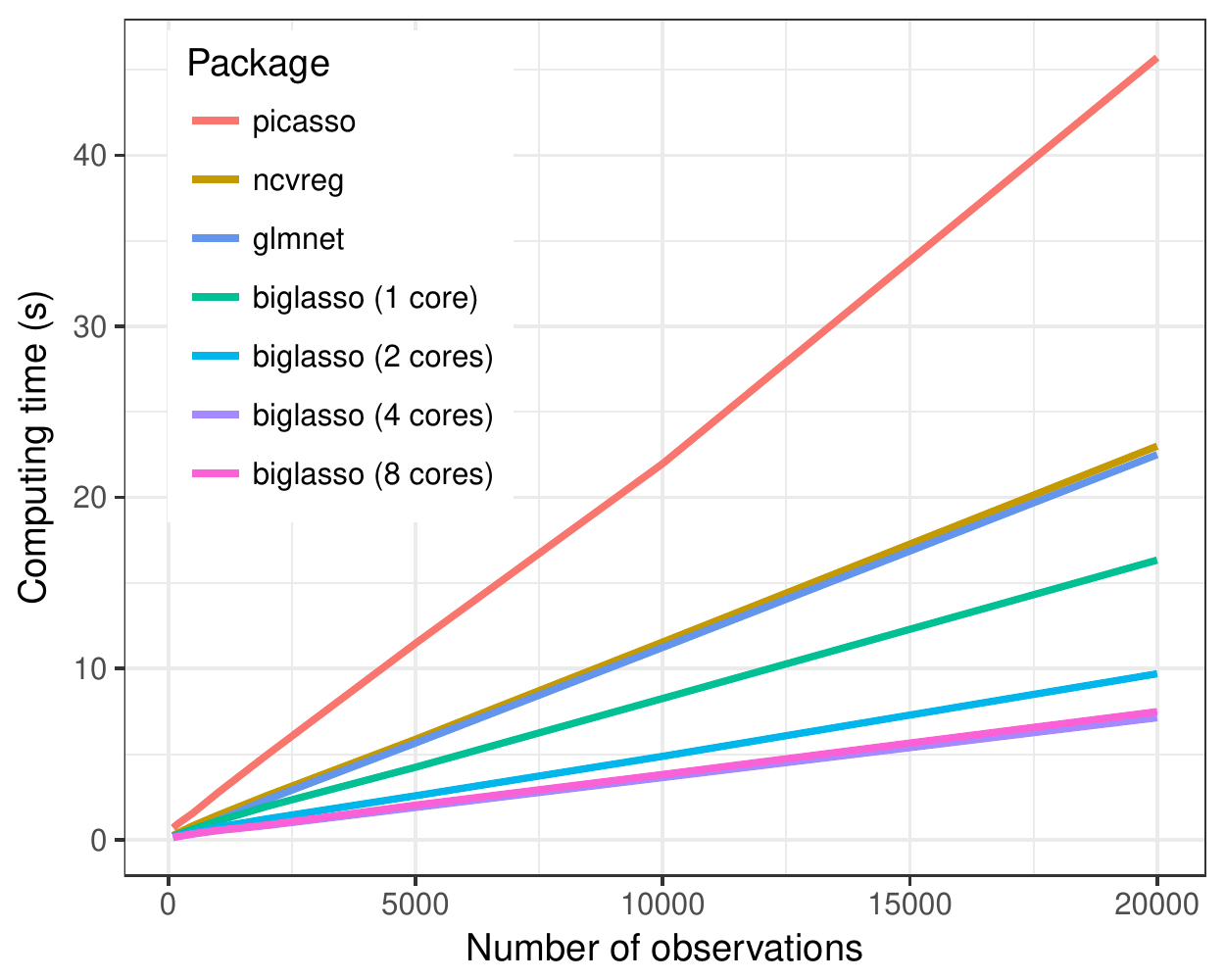}
  \caption{Varying $n$, $p = 10,000$.}
\end{subfigure}
\caption{Mean computing time (in seconds) of solving the lasso-penalized logistic regression over a sequence 100 $\lambda$ values as a function of $p$ (Left) and $n$ (Right).}
\label{fig_simu_res_log}
\end{figure}

Figure~\ref{fig_simu_res_log} summarizes the mean computing times of solving the lasso-penalized logistic regression over a sequence of 100 values of $\lambda$ by the four packages. In all the settings, \pkg{biglasso} (1 core) is around 1.5x faster than \pkg{glmnet} and \pkg{ncvreg}, and more than 3x faster than \pkg{picasso}. Parallel computing with 4 cores using \pkg{biglasso} reduces the computing time by half.

\subsubsection{Real data}

We also compare the computing time of \pkg{biglasso} with other packages for fitting lasso-penalized logistic regression based on four real data sets: (1) Subset of Gisette data set; (2) P53 mutants data set; (3) Subset of NEWS20 data set; (4) Subset of RCV1 text categorization data set. The P53 data set can be found on the UCI Machine Learning Repository website\footnote{\url{https://archive.ics.uci.edu/ml/datasets/p53+Mutants}}~\citep{Lichman:2013}. The other three data sets are obtained from the LIBSVM data repository site.\footnote{\url{https://www.csie.ntu.edu.tw/~cjlin/libsvmtools/datasets/binary.html}}

Table~\ref{tab_real_res_log} presents the dimensions of the data sets and the mean computing times. Again, \pkg{biglasso} outperforms all other packages in terms of computing time in all the real data cases. In particular, It's significantly faster than \pkg{picasso} with the speedup ranging from 2 to 5.5 times (for P53 data and RCV1 data, respectively). On the other hand, compared to \pkg{glmnet} or \pkg{ncvreg}, \pkg{biglasso} doesn't provide as much improvement in speed as in the linear regression case. The main reason is that safe rules for logistic regression do not work as well -- they are more computationally expensive and less powerful in discarding inactive features -- as safe rules for linear regression.

%The size of the feature matrices and the average computing times are summarized in Table \ref{tab_real_res}. Again, \pkg{biglasso} provides 2x to 3.8x speedup compared to  \pkg{glmnet}, and 2x to 4.6x speedup compared to \pkg{picasso}.

\begin{table}[H] 
\centering
%\begin{adjustbox}{max width=\linewidth, center}
\begin{tabular}{r|cccc}
\toprule
%\hline
 & Gisette & P53 & NEWS20 & RCV1 \\
Package & $n = 5,000$ & $n=16,592$ & $n=2,500$ & $n=5,000$ \\
 & $p = 5,000$ & $p=5,408$ & $p=96,202$ & $p=47,236$ \\
\midrule
picasso & 6.15 (0.03) & 19.49 (0.06) & 68.92 (8.17) & 53.23 (0.13)\\
ncvreg & 5.50 (0.03) & 10.22 (0.02) & 38.92 (0.56) & 19.68 (0.07) \\
glmnet & 3.10 (0.02) & 10.39 (0.01) &  25.00 (0.16) & 14.51 (0.04) \\
biglasso & 2.02 (0.01) & 9.47 (0.02) &  18.84 (0.22) & 9.72 (0.04) \\
\bottomrule
\end{tabular}
%\end{adjustbox}
\caption{Mean (SE) computing time (in seconds) for solving the lasso-penalized logistic regression along a sequence of 100 $\lambda$ values on real data sets.}
\label{tab_real_res_log}
\end{table}

%% results with parallel computing
%\begin{table}[H] 
%\centering
%%\begin{adjustbox}{max width=\linewidth, center}
%\begin{tabular}{r|cccc}
%\toprule
%%\hline
% & Gisette & P53 & NEWS20 & RCV1 \\
%Package & $n = 6,000$ & $n=16,592$ & $n=2,500$ & $n=5,000$ \\
% & $p = 5,000$ & $p=5,408$ & $p=96,202$ & $p=47,236$ \\
%\midrule
%picasso & 6.15 (0.03) & 19.49 (0.06) & 68.92 (8.17) & 53.23 (0.13)\\
%ncvreg & 5.50 (0.03) & 10.22 (0.02) & 38.92 (0.56) & 19.68 (0.07) \\
%glmnet & 3.10 (0.02) & 10.39 (0.01) &  25.00 (0.16) & 14.51 (0.04) \\
%biglasso (1 core) & 2.02 (0.01) & 9.47 (0.02) &  18.84 (0.22) & 9.72 (0.04) \\
%biglasso (2 cores) & 1.27 (0.01) & 6.07 (0.01) & 11.84 (0.52) & 6.06 (0.05)\\
%biglasso (4 cores) & 0.99 (0.02) & 4.61 (0.01) &  9.03 (0.42) & 4.61 (0.05) \\
%biglasso (8 cores) & 1.17 (0.02) & 5.14 (0.01) & 9.68 (0.58) & 4.87 (0.02)\\
%\bottomrule
%\end{tabular}
%%\end{adjustbox}
%\caption{Mean (SE) computing time (in seconds) for solving the lasso-penalized logistic regression along a sequence of 100 $\lambda$ values on real data sets.}
%\label{tab_real_res_log}
%\end{table}

\subsection{Validation}

To validate the numerical accuracy of our implementation, we contrast the model fitting results from \pkg{biglasso} to those from \pkg{glmnet} based on the following relative difference criterion:
\begin{equation*}
RD(\lambda) = \frac{\hat{Q}(\hat{\bb}^B; \lambda) - \hat{Q}(\hat{\bb}^G; \lambda)}{\hat{Q}(\hat{\bb}^G; \lambda)},
\end{equation*}
where $\hat{\bb}^B$ and $\hat{\bb}^G$ denote the \pkg{biglasso} and \pkg{glmnet} solutions, respectively.  Four real data sets are considered, including MNIST and GWAS for linear regression, and P53 and NEWS20 for logistic regression. For the GWAS and P53 data sets, we obtain 100 $RD$ values, one of each value of $\lam$ along the regularization path.  For the MNIST and NEWS20 data sets, we obtained solutions for 20 different response vectors, each with a path of 100 $\lam$ values, resulting in 2,000 $RD$ values.

%\begin{itemize}
%\item $\max_{j, l} |\hat{\beta}_j^G(\lambda_l) - \hat{\beta}_j^B(\lambda_l)|$: the maximum absolute difference between coefficient estimates produced by \pkg{glmnet} ($\hat{\beta}_j^G$) and those by \pkg{biglasso} ($\hat{\beta}_j^B$) over the entire solution path.
%\item $RD(\lambda) = \frac{\hat{Q}(\hat{\bb}^B; \lambda) - \hat{Q}(\hat{\bb}^G; \lambda)}{\hat{Q}(\hat{\bb}^G; \lambda)}$: the difference of the objective values obtained by \pkg{biglasso} relative to those by \pkg{glmnet} at each of the 100 values of $\lambda$.
%\end{itemize}

Table~\ref{tab_real_valid} presents the summary statistics of $RD(\lambda)$ for the 4 real data sets. For both linear and logistic regression cases, all values of $RD(\lambda)$ values are extremely close to zero, demonstrating that \pkg{biglasso} and \pkg{glmnet} converge to solutions with virtually identical values of the objective function.

\begin{table}[h]
\centering
%\begin{adjustbox}{max width=\linewidth, center}
  \begin{tabular}{r|rr|rr}
    \toprule
    \multirow{2}{*}{Statistic} &
      \multicolumn{2}{c|}{Linear regression} &
      \multicolumn{2}{c}{Logistic regression} \\
    	  & MNIST & GWAS & P53 & NEWS20 \\
      \midrule
    Minimum           & -7.7e-3 & -3.9e-4 & -6.4e-3 & -1.6e-3  \\
    $1^{st}$ Quantile & -1.6e-3 & -2.7e-5 &  1.7e-5 & -2.2e-4  \\
    Median            & -9.5e-4 &  1.6e-4 &  2.0e-4 & -1.1e-4  \\
    Mean              & -1.1e-3 &  8.3e-4 &  2.2e-4 & -1.2e-4  \\
    $3^{rd}$ Quantile & -1.3e-4 &  1.3e-3 &  7.7e-4 &  1.0e-10 \\
    Maximum           &  4.2e-3 &  4.2e-3 &  2.0e-4 &  2.2e-3  \\
    \bottomrule
  \end{tabular}
%\end{adjustbox}
\caption{Summary statistics of $RD(\lambda)$ based on real data sets.}
\label{tab_real_valid}
\end{table}

\section{Data analysis example} \label{sect:example}

In this section, we illustrate the usage of \pkg{biglasso} with a real data set \code{colon} included in \pkg{biglasso}. The \code{colon} data contains contains expression measurements of 2,000 genes for 62 samples from patients who underwent a biopsy for colon cancer. There are 40 samples from positive biopsies (tumor samples) and 22 from negative biopsies (normal samples). The goal is to identify genes that are predictive of colon cancer.

\pkg{biglasso} package has two main model-fitting R functions as below. Detailed syntax of the two functions can be found in the package reference manual.\footnote{\url{https://cran.r-project.org/web/packages/biglasso/biglasso.pdf}}
\begin{itemize}
\item \code{biglasso}: used for a single model fitting.
\item \code{cv.biglasso}: used for performing cross-validation and selecting parameter $\lambda$.
\end{itemize}

We first load the data: \code{X} is the 62-by-2000 raw data matrix, and \code{y} is the response vector with 1 indicating tumor sample and 0 indicating normal sample.

\begin{CodeInput}
library(biglasso)
data(colon)
X <- colon$X
y <- colon$y
dim(X)
X[1:5, 1:5]
y
\end{CodeInput}

The output of the R snippet above is as follows.
\begin{CodeOutput}
> dim(X)
[1]   62 2000
> X[1:5, 1:5]
  Hsa.3004 Hsa.13491 Hsa.13491.1 Hsa.37254 Hsa.541
t  8589.42   5468.24     4263.41   4064.94 1997.89
n  9164.25   6719.53     4883.45   3718.16 2015.22
t  3825.71   6970.36     5369.97   4705.65 1166.55
n  6246.45   7823.53     5955.84   3975.56 2002.61
t  3230.33   3694.45     3400.74   3463.59 2181.42
> y
 [1] 1 0 1 0 1 0 1 0 1 0 1 0 1 0 1 0 1 0 1 0 1 0 1 0 1 1 1 1 1 1 1 1 1 1 1 1 1
[38] 1 0 1 1 0 0 1 1 1 1 0 1 0 0 1 1 0 0 1 1 1 1 0 1 0
\end{CodeOutput}

\subsection{Set up the design matrix}

It's important to note that \pkg{biglasso} requires that the design matrix \code{X} must be a \code{big.matrix} object - an external pointer to the data. This can be done in two ways:
\begin{itemize}
\item If the size of \code{X} is small, as in this case, a \code{big.matrix} object can be created via:
\begin{CodeInput}
X.bm <- as.big.matrix(X)
\end{CodeInput}
\code{X.bm} is a pointer to the data matrix, as shown in the following output.
\begin{CodeOutput}
> str(X.bm)
Formal class 'big.matrix' [package "bigmemory"] with 1 slot
  ..@ address:<externalptr> 
> dim(X.bm)
[1]   62 2000
> X.bm[1:5, 1:5]
  Hsa.3004 Hsa.13491 Hsa.13491.1 Hsa.37254 Hsa.541
t  8589.42   5468.24     4263.41   4064.94 1997.89
n  9164.25   6719.53     4883.45   3718.16 2015.22
t  3825.71   6970.36     5369.97   4705.65 1166.55
n  6246.45   7823.53     5955.84   3975.56 2002.61
t  3230.33   3694.45     3400.74   3463.59 2181.42
\end{CodeOutput}

\item If the size of the data is large, the user must create a file-backed \code{big.matrix} object via the utility function \code{setupX} in \pkg{biglasso}. Specifically, \code{setupX} reads the massive data stored on disk, and creates memory-mapped files for that data set. Section~\ref{sect:application} demonstrates this case. A detailed example can also be found in the package vignettes.\footnote{\url{https://cran.r-project.org/web/packages/biglasso/vignettes/biglasso.pdf}}

\end{itemize}

\subsection{Single fit and cross-validation} \label{sect:single-fit}

After the setup, we can now fit a lasso-penalized logistic regression model.
\begin{CodeInput}
fit <- biglasso(X.bm, y, family = "binomial") 
\end{CodeInput}
By default, our proposed screening rule ``SSR-Slores'' is used to accelerate the model fitting.

The output object \code{fit} is a list of model fitting results, including the sparse matrix \code{beta}. Each column of \code{beta} corresponds to the estimated coefficient vector at one of the 100 values of $\lambda$.

%\begin{CodeOutput}
%> str(fit)
%List of 17
% $ beta           :Formal class 'dgCMatrix' [package "Matrix"] with 6 slots
%  .. ..@ i       : int [1:1464] 0 0 249 0 249 0 249 0 249 0 ...
%  .. ..@ p       : int [1:101] 0 1 3 5 7 9 11 13 15 17 ...
%  .. ..@ Dim     : int [1:2] 2001 100
%  .. ..@ Dimnames:List of 2
%  .. .. ..$ : chr [1:2001] "(Intercept)" "Hsa.3004" "Hsa.13491" "Hsa.13491.1" ...
%  .. .. ..$ : chr [1:100] "0.3022" "0.2932" "0.2844" "0.276" ...
%  .. ..@ x       : num [1:1464] 5.98e-01 6.34e-01 -3.02e-05 6.70e-01 -5.91e-05 ...
%  .. ..@ factors : list()
% $ iter           : int [1:100] 0 3 2 2 2 2 2 2 2 2 ...
% $ lambda         : num [1:100] 0.302 0.293 0.284 0.276 0.268 ...
% $ penalty        : chr "lasso"
% $ family         : chr "binomial"
% $ alpha          : num 1
% $ loss           : num [1:100] 40.3 39.6 38.9 38.3 37.7 ...
% $ penalty.factor : num [1:2000] 1 1 1 1 1 1 1 1 1 1 ...
% $ n              : int 62
% $ center         : num [1:2000] 7016 4967 4095 3988 2937 ...
% $ scale          : num [1:2000] 3068 2171 1803 2003 1346 ...
% $ y              : num [1:62] 1 0 1 0 1 0 1 0 1 0 ...
% $ screen         : chr "SSR-Slores"
% $ col.idx        : num [1:2000] 1 2 3 4 5 6 7 8 9 10 ...
% $ rejections     : int [1:100] 2000 1998 1998 1997 1996 1996 1996 1996 1996 1995 ...
% $ safe_rejections: int [1:100] 2000 1508 1416 1328 1231 1124 1039 940 848 756 ...
% $ time           : num 0.034
% - attr(*, "class")= chr [1:2] "biglasso" "ncvreg"
%\end{CodeOutput}

%\subsection{Cross validation}
In practice, cross-validation is typically conducted to select $\lambda$ and hence the model with the best prediction accuracy. The following code snippet conducts a 10-fold (default) cross-validation using parallel computing with 4 cores.
\begin{CodeInput}
cvfit <- cv.biglasso(X.bm, y, family = "binomial", seed = 1234, ncores = 4)
par(mfrow = c(2, 2), mar = c(3.5, 3.5, 3, 1), mgp = c(2.5, 0.5, 0))
plot(cvfit, type = "all")
\end{CodeInput}

\begin{figure}[ht]
\centering
\includegraphics[scale=0.4]{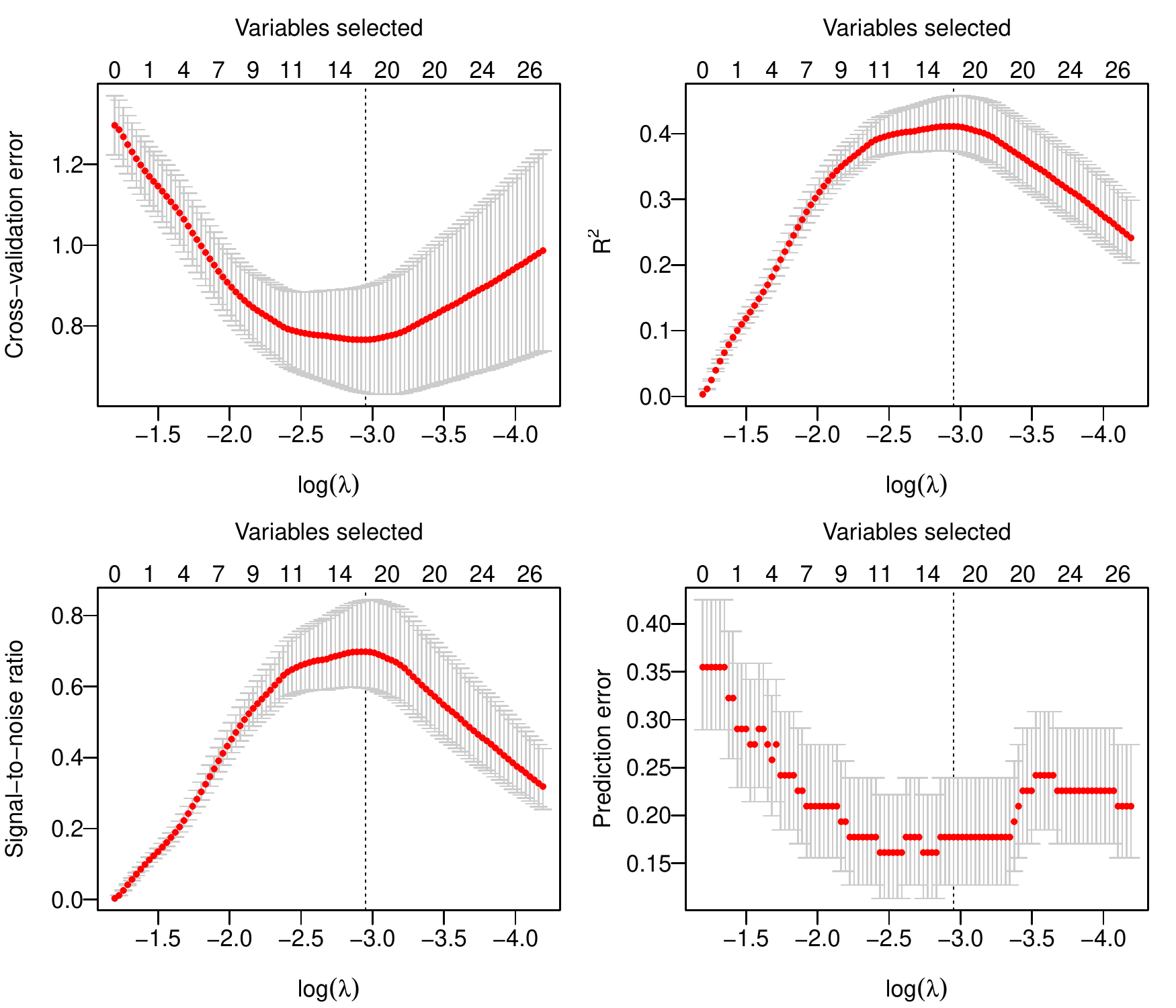}
\caption{The cross-validation curves with standard error bars.}
\label{data_example_cv}
\end{figure}

Figure~\ref{data_example_cv} displays the cross-validation curves with standard error bars. The vertical, dahsed, red line indicates the $\lambda$ value corresponding to the minimum cross-validation error.

\begin{figure}[ht]
\centering
\includegraphics[scale=0.4]{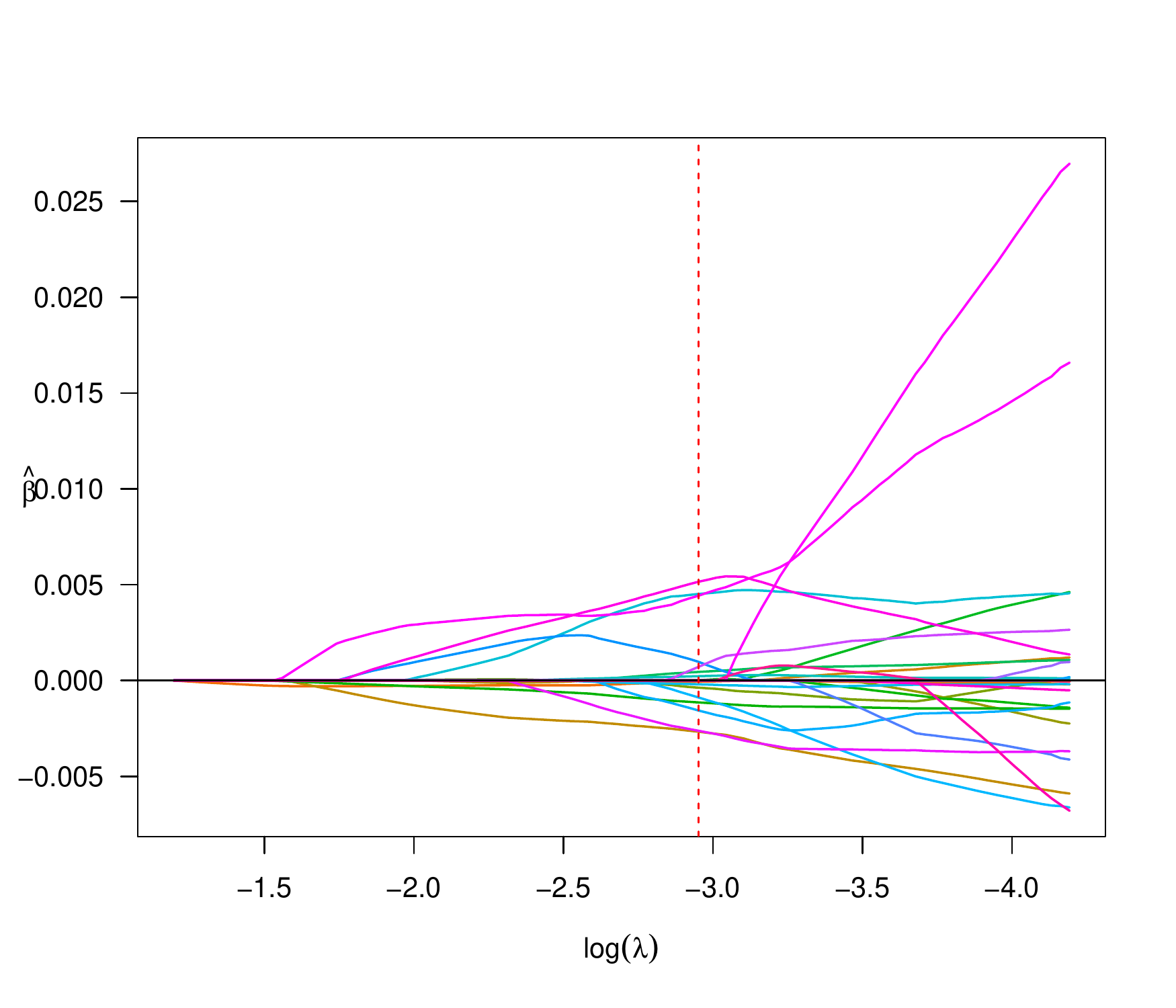}
\caption{The solution path of lasso-penalized logistic regression model for the \code{colon} data.}
\label{data_example}
\end{figure}

Similar to \pkg{glmnet} and other packages, \pkg{biglasso} provides \code{coef}, \code{predict}, and \code{plot} methods for both \code{biglasso} and \code{cv.biglasso} objects.  Furthermore, \code{cv.biglasso} objects contain the \code{biglasso} fit to the full data set, so one can extract the fitted coefficients, make predictions using it, etc., without ever calling \code{biglasso} directly.  For example, the following code displays the full lasso solution path, with a red dashed line indicating the selected $\lambda$ (Figure~\ref{data_example}).

\begin{CodeInput}
plot(cvfit$fit)
abline(v = log(cvfit$lambda.min), col = 2, lty = 2)
\end{CodeInput}

The coefficient estimates at the selected $\lambda$ can be extracted via:
\code{coef}:
\begin{CodeInput}
coefs <- as.matrix(coef(cvfit))
\end{CodeInput}
Here we output only nonzero coefficients:
\begin{CodeOutput}
> coefs[coefs != 0, ]
  (Intercept)      Hsa.8147     Hsa.36689     Hsa.42949     Hsa.22762 
 7.556421e-01 -6.722901e-05 -2.670110e-03 -3.722229e-04  1.698915e-05 
    Hsa.692.2     Hsa.31801      Hsa.3016      Hsa.5392      Hsa.1832 
-1.142052e-03  4.491547e-04  2.265276e-04  4.518250e-03 -1.993107e-04 
    Hsa.12241     Hsa.44244      Hsa.2928     Hsa.41159     Hsa.33268 
-8.824701e-04 -1.565108e-03  9.760147e-04  7.131923e-04 -2.622034e-03 
     Hsa.6814      Hsa.1660 
 4.426423e-03  5.156006e-03
\end{CodeOutput}

The \code{predict} method, in addition to providing predictions for a feature matrix \code{X}, has several options to extract different quantities from the fitted model, such as the number and identity of the nonzero coefficients:
\begin{CodeOutput}
> as.vector(predict(cvfit, X = X.bm, type = "class"))
 [1] 1 0 1 0 1 0 1 0 1 0 1 0 1 0 1 0 1 0 1 0 1 0 1 0 1 1 1 1 1 1 1 1 1
[34] 1 1 1 1 1 0 1 1 0 0 1 1 1 1 0 1 0 1 1 1 0 1 0 1 1 1 0 1 0
> predict(cvfit, type = "nvars")
0.0522 
    16 
> predict(cvfit, type = "vars")
 Hsa.8147 Hsa.36689 Hsa.42949 Hsa.22762 Hsa.692.2 Hsa.31801  Hsa.3016 
      249       377       617       639       765      1024      1325 
 Hsa.5392  Hsa.1832 Hsa.12241 Hsa.44244  Hsa.2928 Hsa.41159 Hsa.33268 
     1346      1423      1482      1504      1582      1641      1644 
 Hsa.6814  Hsa.1660 
     1772      1870 
\end{CodeOutput}

In addition, the \code{summary} method can be applied to a \code{cv.biglasso} object to extract useful cross-validation results:
\begin{CodeOutput}
> summary(cvfit)
lasso-penalized logistic regression with n=62, p=2000
At minimum cross-validation error (lambda=0.0522):
-------------------------------------------------
  Nonzero coefficients: 16
  Cross-validation error (deviance): 0.77
  R-squared: 0.41
  Signal-to-noise ratio: 0.70
  Prediction error: 0.177
\end{CodeOutput}

\section{Application: Big Data case} \label{sect:application}

Perhaps the most important feature of \pkg{biglasso} is its capability of out-of-core computing. To demonstrate this, we use it to analyze data from a genome-wide association study of 2898 infants (1911 controls and 987 premature cases). The goal is to identify genetic risk factors for premature birth. After preprocessing, 1,339,511 SNP (single-nucleotide polymorphism) measurements are used as features. The size of the resulting feature matrix is over 31 GB data, which is nearly 2x larger than the installed 16 GB RAM.

In this Big Data case, the data is stored in an external file on the disk. To use \pkg{biglasso}, memory-mapped files are first created via the following command.

\begin{CodeInput}
X <- setupX(filename = "gwas_data.raw")
\end{CodeInput}

This command creates two files in the current working directory: (1) a memory-mapped file cache of the data, \code{gwas_data.bin}; and (2) a descriptor file, \code{gwas_data.desc}, that contains the backingfile description.

Note that this setup process takes a while if the data file is large (approximately 30 minutes in this case). However, this only needs to be done once, during data processing.  Once the cache and descriptor files are generated, all future analyses using \pkg{biglasso} can use the \code{X} object.  In particular, should one close R and open a new R session at a later date, \code{X} can be seamlessly retrieved by attaching its descriptor file as if it were already loaded into the main memory:

\begin{CodeInput}
X <- attach.big.matrix("gwas_data.desc")
\end{CodeInput}

The object \code{X} returned from \code{setupX} or \code{attach.big.matrix} is a \code{big.matrix} object that is ready to be used for model fitting. Details about \code{big.matrix} and its related functions such as \code{attach.big.matrix} can be found in the reference manual of \pkg{bigmemory} package.\footnote{\url{https://cran.r-project.org/web/packages/bigmemory/bigmemory.pdf}}

Note that the object \code{X} that we have created is a \code{big.matrix} object and is therefore stored on disk, not in RAM, but can be accessed as if it were a regular \proglang{R} object:
\begin{CodeOutput}
> str(X)
Formal class 'big.matrix' [package "bigmemory"] with 1 slot
  ..@ address:<externalptr> 
> dim(X)
[1]    2898 1339511  
> X[1:10, 1:10]
      [,1] [,2] [,3] [,4] [,5] [,6] [,7] [,8] [,9] [,10]
 [1,]    1    1    0    0    1    1    1    0    1     1
 [2,]    1    0    1    1    1    1    0    0    1     1
 [3,]    1    0    1    1    1    1    0    0    1     1
 [4,]    1    1    1    0    1    1    0    1    1     1
 [5,]    1    1    0    1    1    1    0    1    1     1
 [6,]    1    2    0    1    1    0    1    2    0     0
 [7,]    1    2    0    0    1    0    1    2    1     0
 [8,]    1    1    0    0    1    1    1    1    0     1
 [9,]    1    0    1    1    1    1    0    0    0     1
[10,]    1    1    1    0    1    1    1    1    0     1
> table(y)

   0    1 
1911  987 
\end{CodeOutput}

Here we fit both sparse linear and logistic regression models with the lasso penalty over the entire path of 100 $\lambda$ values equally spaced on the scale of $\lambda / \lambda_{\max}$. The ratio of $\lambda_{\min} / \lambda_{\max}$ is set to be 0.1 for both models. Parallel computation with 4 cores is applied:

\begin{CodeInput}
fit <- biglasso(X, y, ncores = 4)
fit <- biglasso(X, y, family = "binomial", ncores = 4)
\end{CodeInput}

\begin{figure}[ht]
\centering
\includegraphics[scale=0.4]{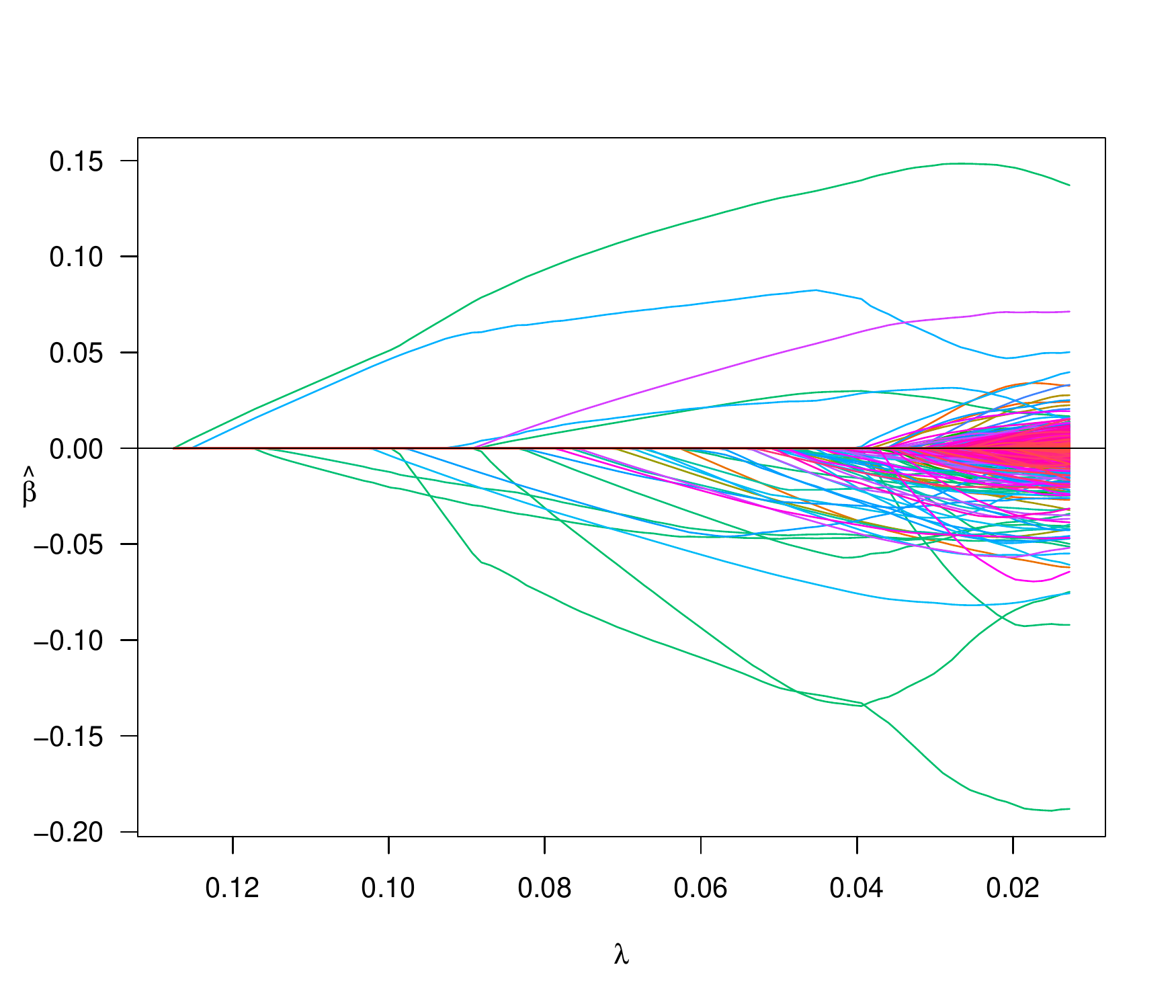}
\caption{The solution path of the sparse linear regression model for the 31 GB GWAS data.}
\label{big_data_linear}
\end{figure}

The above code, which solves the full lasso path for a 31 GB feature matrix, required 94 minutes for the linear regression fit and 146 minutes for the logistic regression fit on an ordinary laptop with 16 GB RAM installed.  Figure~\ref{big_data_linear} depicts the lasso solution path for the sparse linear regression model. The following code extracts the nonzero coefficient estimates and the number of selected variables of the lasso model when $\lambda = 0.06$:

\begin{CodeOutput}
> coefs <- as.matrix(coef(fit, lambda = 0.06))
> coefs[coefs != 0, ]
 (Intercept)       V78255      V289696      V602992      V602997      V602999 
 1.510095538 -0.004741494 -0.014724867  0.119765071 -0.093618971  0.020834021 
     V603003      V603034      V603042      V603065      V603067      V646528 
-0.109134982 -0.046235902 -0.035036718 -0.042304267 -0.002034167 -0.019205575 
     V646544      V773260      V785150      V806320      V828475      V828488 
-0.009025915 -0.012615210 -0.055444964 -0.011623245  0.075476347  0.021059097 
     V877563      V877564      V877624     V1000488     V1001897     V1136725 
-0.044588035 -0.001164987 -0.020683459 -0.020375492  0.038315644 -0.021952476 
> predict(fit, lambda = 0.06, type = "nvars")
0.06 
  23 
\end{CodeOutput}

%% \begin{table}[h]
%% \centering
%% %\begin{adjustbox}{max width=\linewidth, center}
%%   \begin{tabular}{r|c|c}
%%     \toprule
%%     & Linear regression & Logistic regression \\
%%       \midrule
%%     Time & 94 & 146 \\
%%     \bottomrule
%%   \end{tabular}
%% %\end{adjustbox}
%% \caption{The total computing time (in minutes) of solving the sparse linear and logistic regression models with the lasso penalty along a sequence of 100 $\lambda$ values for 31 GB data on a laptop with 16 GB RAM.}
%% \label{big_data_time}
%% \end{table}

\section{Conclusion} \label{sect:conclusion}

We developed a memory- and computation-efficient R package \pkg{biglasso} to extend lasso model fitting to Big Data.
The package provides functions for fitting regularized linear and logistic regression models with both lasso and elastic net penalties.
Equipped with the memory-mapping technique and more efficient screening rules, \pkg{biglasso} is not only is 1.5x to 4x times faster than existing packages, but consumes far less memory and, critically, enables users to fit lasso models involving data sets that are too large to be loaded into memory.

%Since its first release in March 2016, \pkg{biglasso} has drawn increasing attention from R users. The number of its CRAN downloads is 4,016 as of January 19, 2017.

%% It's important to note that the idea of ``hybrid safe-strong rule'' (i.e., combining SSR with a \textit{simple} safe rule) is a rather general screening paradigm that could apply to and thus speed up a broad range of sparse learning problems. For example, we have developed a new screening rule, called SSR-Slores, for lasso-penalized logistic regression. We have also extended SSR-BEDPP to sparse linear regression with elastic-net and group lasso penalties.

\section*{Appendix} \label{supple}

Using the big data set in Section~\ref{sect:application}, we also conduct experiments to investigate the time savings by our newly proposed rule, SSR-BEDPP, compared to SSR.

Here we solve lasso-penalized linear regression over the entire path of 100 $\lambda$ values equally spaced on the scale of $\lambda / \lambda_{\max}$. We consider two cases in terms of $\lambda_{\min}$: (1) $\lambda_{\min} = 0.1 \lambda_{\max}$ as before; and (2) $\lambda_{\min} = 0.5 \lambda_{\max}$, as in practice there is typically less interest in lower values of $\lam$ for very high-dimensional data such as this.

Since other packages cannot handle this data-larger-than-RAM case, we compare the performance of screening rules SSR and SSR-BEDPP based on \pkg{biglasso}. Table \ref{tab_big_data} summarizes the total computing time with the two rules under two cases with and without parallel computing. Clearly our proposed rule SSR-BEDPP is much more efficient in both cases. In particular, SSR-BEDPP is nearly 2.8x faster than SSR in Case 2.

\begin{table}[h]
\centering
%\begin{adjustbox}{max width=\linewidth, center}
  \begin{tabular}{r|rr|rr}
    \toprule
    \multirow{2}{*}{Rule} &
      \multicolumn{2}{c|}{$\lambda_{\min} / \lambda_{\max} = 0.1$} &
      \multicolumn{2}{c}{$\lambda_{\min} / \lambda_{\max} = 0.5$} \\
    	  & 1 core & 4 cores & 1 core & 4 cores \\
      \midrule
    SSR & 285 & 143 & 286 & 141 \\
    SSR-BEDPP & 189 & 94 & 103 & 51 \\
    \bottomrule
  \end{tabular}
%\end{adjustbox}
\caption{The total computing time (in minutes) of solving the lasso-penalized linear regression along a sequence of 100 $\lambda$ values for 31 GB data on a laptop with 16 GB RAM.}
\label{tab_big_data}
\end{table}

\end{document}